\documentclass[preprint,showpacs,floatfix] {revtex4} 
\usepackage{graphicx}
\usepackage{subfigure}
\usepackage{longtable}
\newcommand{\rvec}{\mathrm{\mathbf{r}}}
\newcommand{\rpvec}{\mathrm{\mathbf{r'}}}
\begin{document}
\title{Studies on the hollow states of atomic lithium by a density functional
approach}
\author{Amlan K. Roy}
\affiliation{Department of Chemistry, University of New Brunswick, Fredericton,
NB, E3B 6E2, Canada}
\email{akroy@unb.ca}
\begin{abstract}
Density functional calculations are performed for twelve $2l2l'$n$l''$ (n$\geq$2)
triply excited hollow resonance series of Li, {\em viz.,} 2s$^2$ns $^2$S$^e$, 2s$^2$np 
$^2$P$^o$, 2s$^2$nd $^2$D$^e$, 2p$^2$ns $^2$D$^e$,$^4$P$^e$, 2s2pns $^4$P$^o$, 2s2pnp 
$^4$D$^e$, 2p$^2$np $^2$F$^o$,$^4$D$^o$, 2p$^2$nd $^2$G$^e$, $^4$F$^e$ and 2s2pnd 
$^4$F$^o$, covering a total of about 270 low-, moderately high- and high-lying states, 
with n as high as up to 25. The work-function-based exchange potential and the nonlinear 
gradient plus Laplacian included Lee-Yang-Parr correlation energy functional is
used. The relevant Kohn-Sham-type equation is solved numerically using the generalized 
pseudospectral method offering nonuniform, optimal spatial discretization to obtain
the orbitals and densities. The present single determinantal approach yields fairly 
accurate results for the nonrelativistic energies, excitation energies as well as the 
radial densities and other expectation values. Except for the one state, the discrepancy
in the calculated state energies remains well within 0.98\%, whereas the excitation 
energies deviate by 0.02--0.58\% compared to the available experimental and other
theoretical results. Additionally companion calculations are also presented for the 
37 $3l3l'$n$l''$ (n$\geq$3) doubly hollow states (seven are n=3 intrashell type) of Li
with both K and L shells empty (up to n=6) in the photon energy range 175.63--180.51 
eV, with varying symmetries and multiplicities. Our calculation shows good agreement 
with the recent literature data for the only two such doubly hollow states reported so
far, {\em viz.,} 3s$^2$3p $^2$P$^o$ and 3s3p$^2$ $^4$P$^e$. The resonance series are 
found to be inextricably entangled to each other, leading to complicated behavior in 
their positions. Many new states are reported here for the first time. This 
provides a simple, efficient and general scheme for the accurate calculation of these
and other multiply excited Rydberg series of many-electron atomic systems within 
density functional theory.
\end{abstract}
\maketitle

\section{Introduction}
Triply excited states of atomic lithium represent the prototypical case of a highly
correlated, multi-excited three-electron system in the presence of a nucleus, and thus 
typify the four-body Coulombic problem. Since all the three electrons reside in 
higher shells leaving the K shell empty, these are often referred to as the \emph 
{hollow} states. As the one-step photogeneration of such a state requires coherent 
excitation of all the three electrons, these are usually more difficult to produce 
from the ground state by single photon absorption or electron impact excitation. 
These, in addition to their proximity to more than one thresholds coupled with the 
presence of an infinite number of open channels associated with these resonances, 
have offered considerable challenges to both experimentalists as well as theoreticians.  

The recent advances in third-generation, extreme-UV synchrotron radiation sources as 
well as the availability of sophisticated and powerful quantum mechanical 
methodologies have stimulated an overwhelming amount of works in the last decade that
shed significant light on the understanding of such systems with greater accuracy. 
Ever since the first observation of $2l2l'2l''$ states of Li in a beam-foil 
experiment [1], numerous attempts [2-4] have been made over the years to use this 
technique which identified bound states such as the 2p$^3$ $^4$S$^o$ besides many 
autoionizing states. Later, photoabsorption spectroscopy [5] using a dual laser 
plasma technique reported measurement of the lowest 2s$^2$2p $^2$P$^o$ resonance in 
Li. Subsequently a large number of higher resonances were found and tentatively 
classified in a wider energy range (140--165 eV) [6-8]. First high-precision 
photoelectron spectroscopic determination [9] of the partial photoionization cross 
sections generated tremendous interest to measure the resonance positions, widths of
both even- and odd-parity hollow states having varied symmetries [10-14] with better
resolution over wider energy regions. Some Rydberg series [15] as well as the states
with both K and L shell vacancies [8,13] have also been identified lately.

Parallel to the experimental progresses, a substantial amount of theoretical works 
have also been done over the years for reliable, accurate prediction of energies, 
widths and lifetimes of these multi-excited states, varying in their complexity and 
accuracy. One of the earliest such calculations for Li-isoelectronic series was due 
to [16] employing a truncated diagonalization method (TDM), based on the hydrogenic 
orbitals. Later other formalisms such as the 1/Z expansion method [17], many-body 
perturbation theory [18] were also used to investigate these states. Some other 
recent theoretical works include the state-specific theory [19-21], the configuration 
interaction (CI) type approach [22], joint saddle point (SP) and complex coordinate
rotation (CCR) method [23-25], the space partition and stabilization procedure [26],
a density functional formalism [27], several variants of the R-matrix theory [28-31]
as well as the TDM [32,33], the hyperspherical coordinate approach [34,35], etc. 
Although the most natural and commonest choice in these works was neutral atom (Li), 
other atomic systems such as the negative ions (e.g., He$^-$), positive ions like 
Li-isoelectronic series or N$^{2+}$ [22] have also received considerable attention. 
While majority of the available formalisms involve traditional wave-function-based 
calculations requiring large basis sets and often involving mixing of continuum 
states, a density-based formalism for triply excited states was proposed in [27]. In
this approach a Kohn-Sham (KS) type differential equation was solved to obtain the 
self-consistent set of orbitals and densities using the nonvariational local 
work-function based exchange potential [36] in conjunction with the nonlinear 
gradient- and Laplacian-included correlation energy functional of Lee, Yang and Parr
(LYP) [37]. Fairly good quality results were reported for various closed and open 
shell many-electron atomic excited states including the single, double, triple as 
well as low and moderately high excitations; core and valence states; bound and 
autoionizing states besides the satellite states [38-43,27]. This single-determinantal
formalism was able to reproduce the excited state energies, radial densities, various
expectation values, excitation energies (10--2000 eV) as well as the small energy 
differences (0.03--23.5 eV) quite satisfactorily. However, all the above density 
functional theory (DFT)-based calculations of atomic excited states employed a 
Numerov-type finite difference (FD) scheme to numerically solve the KS type equation.
Recently a generalized pseudospectral (GPS) approach has been utilized to solve it 
in a more accurate and efficient manner [44], as demonstrated for the singly and 
doubly excited high-lying Rydberg states of He as well as the singly excited states
of Li and Be. The singly excited state energies were within 0.01\% of the best 
literature data for He (for other atoms it was less than 0.02\%), while for the 
doubly excited states of He it was well within 3.6\%. The discrepancy in the 
calculated single and double excitation energies for 31 selected states were within
0.009--0.632\% and 0.085--1.600\% respectively. 

The GPS method has been shown to be a very powerful tool for accurate and physically
meaningful results of both static and dynamic properties of Coulombic singular
systems such as atoms and molecules as well as other stronger singularities like
the generalized spiked harmonic oscillators [44-48]. While in [27], results were 
presented for all the eight n=2 intrashell triply excited states of Li isoelectronic 
series (Z=2,3,4,6,8,10) with reasonable accuracy, resonance series including the higher
members were not considered and to the best of our knowledge, no other DFT-based
attempts are known so far for these systems. The purpose of this work is therefore
to explore and extend the regions of validity of the density-based approach to these 
challenging hollow states and also to study the spectra of these resonances in a 
detailed and elaborate way. We thus focus on the energies and the excitation energies
of twelve $2l2l'$n$l''$ (n$\geq$2) odd- and even-parity doublet and quartet resonance
series of Li leading nearly to 270 states (2s$^2$ns $^2$S$^e$, 2s$^2$np $^2$P$^o$, 
2s$^2$nd $^2$D$^e$, 2p$^2$ns $^2$D$^e$,$^4$P$^e$, 2s2pns $^4$P$^o$, 2s2pnp $^4$D$^e$,
2p$^2$np $^2$F$^o$, $^4$D$^o$, 2p$^2$nd $^2$G$^e$, $^4$F$^e$, 2s2pnd $^4$F$^o$), up to
a maximum of n=25. As a further test, additional results are also reported for 37 
$3l3l'$n$l''$ (n$\geq$3) doubly hollow states of Li having vacancies in both the K and
L shells up to n=6, out of which seven are n=3 intrashell type. Comparisons are made
with the experimental and other theoretical results, wherever possible. The article 
is organized as follows: Section II presents a brief summary of the methodology as 
well as the computational aspects. Section III makes a detailed discussion on the 
obtained results and a few conclusions are drawn in section IV. 

\section{Methodology} 

Since the work-function formalism for atomic excited states [38-43,27] as well as 
its GPS implementation [44] has been discussed in detail previously, here we give
only the essential steps in the calculation (atomic units employed throughout the
article, unless otherwise mentioned).

The local exchange potential with which the electrons move is interpreted [36,49] 
physically as the work done to move an electron from infinity to its position $\rvec$
against the electric field $\mbox{\boldmath $\cal{E}$}_x(\rvec)$ arising out of the
Fermi-hole charge distribution, $\rho_x (\rvec,\rpvec)$, and as such is given by the
following line integral,
\begin{equation}
\mathit{v}_x (\rvec) = - \int_{\infty}^{r} 
\mbox{\boldmath $\cal{E}$}_x (\rvec) \cdot \mathrm{d} \mathbf{l}.
\end{equation}
This field $\mbox{\boldmath $\cal{E}$}_x(\rvec)$, being representative of the Pauli 
correlation, as its quantum mechanical source charge distribution is the pair 
correlation density, has the following form,
\begin{equation} \label{eq:eqn1}
\mbox{\boldmath $\cal{E}$}_x(\rvec) = \int \frac 
{\rho_x (\rvec,\rpvec)(\rvec -\rpvec)} {|\rvec-\rpvec|^3} \ \ \mathrm{d}\rvec. 
\end{equation}
This work against $\mbox{\boldmath $\cal{E}$}_x(\rvec)$ can be determined exactly as
the Fermi hole is known explicitly in terms of the single-particle orbitals,
\begin{equation}
\rho_x(\rvec,\rpvec)=-\frac{|\gamma (\rvec,\rpvec)|^2}{2 \rho(\rvec)},
\end{equation}
where  
\begin{equation}
\gamma(\rvec,\rpvec)=\sum_{\mathit{i}} \phi_{\mathit{i}}^\ast(\rvec) 
\phi_{\mathit{i}}(\rpvec).
\end{equation}
denotes the single-particle density matrix spherically averaged over coordinates of 
the electrons of a given orbital angular momentum quantum number. Now within the 
central-field approximation, the orbitals expressed as $\phi_i(\rvec)= R_{nl}(r)\ 
Y_{lm}(\Omega)$ ($Y_{lm}(\Omega)$ denoting the usual spherical harmonics), give the
total electron density as the sum of the occupied orbitals,
\[\rho(\rvec)=\sum_{\mathit{i}} |\phi_{\mathit{i}}(\rvec)|^2. \]
and the spherically averaged radial component of the electric field is simplified as,
\begin{equation}
\mbox{\boldmath $\cal{E}$}_{x,r} (r)=-\frac{1}{4\pi} \int 
\rho_x (\rvec,\rpvec) \ \frac{\partial}{\partial r} \ \ 
\frac{1}{|\rvec -\rpvec|}\ \mathrm{d}\rpvec \mathrm{d} \Omega_r.
\end{equation}
Now assuming that a unique local exchange potential exists for a given excited state,
the following KS-type equation is solved,
\begin{equation}
\left[ -\frac{1}{2} \nabla^2 +\mathit{v}_{es} (\rvec) +\mathit{v}_{xc}(\rvec)
\right] \phi_i(\rvec) = \varepsilon_{\mathit{i}} \phi_{\mathit{i}}(\rvec),
\end{equation}
to obtain the self-consistent set of orbitals $\{\phi_i\}$ and the electron density.
Here $\mathit{v}_{es}(\rvec)$ denotes the usual Hartree electrostatic potential 
consisting of the electron-nuclear attraction and the interelectronic Coulomb repulsion,
\begin{equation}
\mathit{v}_{es}(\rvec)=-\frac{Z}{r} + \int \frac{\rho(\rpvec)}{|\rvec-\rpvec|} 
\mathrm{d}\rpvec
\end{equation}
and $\mathit{v}_{xc}(\rvec)$, the total exchange-correlation (XC) potential is
partitioned as,
\begin{equation}
\mathit{v}_{xc}(\rvec)=\mathit{v}_{x}(\rvec)+\mathit{v}_{c}(\rvec)
\end{equation}
While $\mathit{v}_{x}(\rvec)$ can be accurately determined as outlined above, the 
accurate form of $\mathit{v}_{c}(\rvec)$ valid for a general excited state is unknown 
as yet and therefore requires to be approximated. There are several prescriptions 
available in the literature and the present work uses the widely used LYP functional
[37]. 

Now we present a brief overview of the computational procedure used to solve the
pertinent KS equation through the GPS formalism. One disquieting feature of the 
commonly used equal-spacing FD methods is that because of the existence of Coulomb
singularity at the origin as well as the long range nature of the Coulomb potential,
one requires a large number of grid points to achieve reasonably good accuracy even 
for the ground states. However the GPS method allows nonuniform and optimal spatial
discretization maintaining similar accuracy at both small and large $r$ regions. 
Thus one can work with a much lesser grid points having a denser mesh at small $r$
while a coarser mesh at large $r$. Moreover it also has the attractive feature of
possessing both the simplicity of direct FD or finite element methods and the fast
convergence of the finite basis set methods. 

One of the principal features of this scheme lies in the fact that a function $f(x)$
defined in the interval $x \in [-1,1]$ can be approximated by the polynomial $f_N(x)$
of order N so that,  
\begin{equation}
f(x) \cong f_N(x) = \sum_{j=0}^{N} f(x_j)\ g_j(x),
\end{equation}
and the approximation is \emph {exact} at the \emph {collocation points} $x_j$, i.e.,
\begin{equation}
f_N(x_j) = f(x_j).
\end{equation}
Here we employ the Legendre pseudospectral method using $x_0=-1$, $x_N=1$, where 
$x_j (j=1,\ldots,N-1)$ are obtainable from the roots of the first derivative of the 
Legendre polynomial $P_N(x)$ with respect to $x$, i.e., 
\begin{equation}
P'_N(x_j) = 0.
\end{equation}
The cardinal functions, $g_j(x)$ in Eq.~(9) are given by the following expression,
\begin{equation}
g_j(x) = -\frac{1}{N(N+1)P_N(x_j)}\ \  \frac{(1-x^2)\ P'_N(x)}{x-x_j},
\end{equation}
obeying the unique property $g_j(x_{j'}) = \delta_{j'j}$. Now the semi-infinite 
domain $r \in [0, \infty]$ is mapped into the finite domain $x \in [-1,1]$ by the 
transformation $r=r(x)$. One can make use of the following algebraic nonlinear 
mapping [50,51], 
\begin{equation}
r=r(x)=L\ \ \frac{1+x}{1-x+\alpha},
\end{equation}
where L and $\alpha=2L/r_{max}$ may be termed as the mapping parameters. Now, 
introducing the following relation, 
\begin{equation}
\psi(r(x))=\sqrt{r'(x)} f(x)
\end{equation}
coupled with the symmetrization procedure [50,51] leads to the following coupled set
of equations,  
\begin{widetext}
\begin{equation}
\sum_{j=0}^N \left[ -\frac{1}{2} D^{(2)}_{j'j} + \delta_{j'j} \ v(r(x_j))
+\delta_{j'j}\ v_m(r(x_j))\right] A_j = EA_{j'},\ \ \ \ j=1,\ldots,N-1,
\end{equation}
\end{widetext}
where
\begin{equation}
A_j  = \left[ r'(x_j)\right]^{1/2} \psi(r(x_j))\ \left[ P_N(x_j)\right]^{-1}.
\end{equation}
and the symmetrized second derivative of the cardinal function, $D^{(2)}_{j'j}$ is 
given by,
\begin{equation}
D^{(2)}_{j'j} =  \left[r'(x_{j'}) \right]^{-1} d^{(2)}_{j'j} 
\left[r'(x_j)\right]^{-1}, 
\end{equation}
with
\begin{eqnarray}
d^{(2)}_{j',j} & = & \frac{1}{r'(x)} \ \frac{(N+1)(N+2)} {6(1-x_j)^2} \ 
\frac{1}{r'(x)}, \ \ \ j=j', \nonumber \\
 & & \nonumber \\
& = & \frac{1}{r'(x_{j'})} \ \ \frac{1}{(x_j-x_{j'})^2} \ \frac{1}{r'(x_j)}, 
\ \ \ j\neq j'.
\end{eqnarray}
Note the advantage that this leads to a \emph {symmetric} matrix eigenvalue problem 
which can be readily solved to give accurate eigenvalues and eigenfunctions. 

Within the central-field approximation, taking the exchange potential in Eq.~(1) 
and the LYP correlation functional [37], KS equation (6) is solved numerically in a 
self-consistent manner to obtain the orbitals. These are then employed to construct
the various Slater determinants derived from a particular electronic configuration. 
In general, denoting the electronic energies of determinants and multiplets 
associated with a given electronic configuration by $E(D_i)$ and $E(M_i)$ 
respectively, the Slater's diagonal sum rule [52,38-44,27] is used to calculate the
$E(M_i)$ as,
\begin{equation}
E(M_j)=\sum_i A_{ji} E(D_i).  
\end{equation}
For all the calculations reported in this work, a consistent convergence criteria of
$10^{-6}$ and $10^{-8}$ a.u., are used for the potential and energy respectively, 
whereas a maximum of 500 radial grid points sufficed to achieve convergence.

\section{Results and discussion}
The calculated nonrelativistic term and excitation energies for various hollow states
of Li along with the available experimental and theoretical results are given in 
tables I through VII. In the literature, usually the excitation energies are reported, 
while the individual state energies are given in only few occasions. In a recent study
of the low-lying singly excited states of several open-shell atoms (B, C, O, F, Na, 
Mg, Al, Si, P and Cl) within the work-function formalism [42], it was observed that the 
excitation energies from the exchange-only and the numerical Hartree-Fock (HF) 
calculations showed good agreement with each other. Interestingly however, the two
correlation energy functionals (relatively simpler local Wigner and the nonlocal LYP) 
did not show any considerable improvements in excitation energies, although the 
excited-state energies were improved significantly. Therefore we decided to present 
both these quantities for all the states considered in this work. The state energies 
are given in atomic units, while the excitation energies in eV (1 a.u.=27.2076544 eV) 
are given relative to the accurate ground state of Li calculated from a full core plus 
correlation using the multiconfiguration interaction wave function, i.e., $-7.47805953$ 
a.u. [54], for a consistent comparison with the literature data. Here we
note that the present calculated ground state energy of Li is $-7.4782839$ a.u. The 
independent particle model classification [32,33] is used throughout the article, 
where the six core Li$^+$ n=2 intrashell doubly excited states, {\em viz.,} 2s$^2$ 
$^1$S$^e$, 2s2p $^3$P$^o$, 2p$^2$ $^3$P$^e$, 2p$^2$ $^1$D$^e$, 2s2p $^1$P$^o$ and 
2p$^2$ $^1$S$^e$ are denoted by A, B, C, D, E and F respectively. Only the state 
energies were reported in [33], so their excitation energies quoted in this work for 
comparison were calculated from the same Li ground state as mentioned above. The 
excitation energies in [31] were given with respect to the first excited 1s$^2$ 
$^2$P$^o$ state of Li and thus to put things in proper perspective, we have used 
E[1s$^2$2p $^2$P$^o$$-$1s$^2$2s $^2$S$^e$]=1.848 eV to obtain the respective resonance
positions relative to the Li ground state. 

Now table I tabulates the results for even-parity $\langle$A,ns$\rangle$ $^2$S$^e$ and
odd-parity $\langle$A,np$\rangle$ $^2$P$^o$, n=2--22 hollow states of Li. The 
experimental results for $^2$S$^e$ resonances are yet to be obtained. The lower 
members of the series have been treated theoretically in considerable detail by 
several workers employing a multitude of techniques, e.g., the n=3 state by a 
hyperspherical coordinate method [35], the n=3,4 resonances by a combination of SP and
CCR method [25]. The n=3--9 resonances were also calculated by the R-matrix method 
including a set of core base states for the Li$^+$ target along with an optimization
of the collisional representation [28]. Recently, an eigenphase derivative technique
in conjunction with a quantum defect theory [31] reported the low and high resonances
up to n=22, whereas the same up to n=12 have been studied through the TDM [33]. The
agreement of the present density functional results for both the quantities are seen
to be fairly good; the current state energies lie about 0.36--0.88\% above the TDM 
results [33], whereas the excitation energies are higher by 0.41--0.51\% from those of
[31]. It is noted that the small differences in excitation energies are reproduced
quite nicely within the whole series; a feature which holds true for all the resonance
series dealt in this work. Coming to the $^2$P$^o$ states, these are the most widely
studied hollow resonance series in Li, both experimentally and theoretically, chiefly
due to the fact that the lowest n=2 intrashell Li state is the 2s$^2$2p $^2$P$^o$. 
Photoion yield spectroscopic measurements determined its position at 142.33 eV [7] and
142.35 eV [6] respectively, which has been supported by various theoretical 
calculations, e.g., the combined SP CCR method [23], the complex scaling method having
correlated basis functions constructed from B-splines [53], the R-matrix method [28],
as well as the TDM method [33]. Our calculation gives an excitation energy of 142.385
eV which is only 0.02\% above the experimental results. It may be noted that the HF 
energy of the lowest resonance is $-$2.16954 a.u. [18], which matches almost exactly 
with the present exchange-only result of $-$2.1694 a.u., (0.00014 a.u., deviation) 
illustrating the fact that the exchange contribution is included quite accurately 
within this formalism. Other members of the series with n=3--7 show good agreement with
the CCR calculation [24]. For n=8--11, the term energies are available also through the
TDM calculation [33] and fairly good agreement is noticed in our results with those. 
The present term energies are underestimated by 0.24--0.98\% with respect to the CCR
results [23,24], leading to higher excitation energies (the deviations with respect to 
[28], [29] being 0.05-0.32\% and 0.43--0.46\% for n=2--9 and n=3--10 respectively). No
results could be found for the resonances with n$\ge$12. 

\begingroup
\squeezetable
\begin{table}
\caption {\label{tab:table1}Comparison of the state (in a.u.) and excitation energies
(in eV) relative to the Li ground state [54], of 2s$^2$ns $^2$S$^e$ and 2s$^2$np 
$^2$P$^o$ resonances of Li. PW signifies present work.}
\begin{ruledtabular}
\begin{tabular}{lllllllll}
n   & \multicolumn{4}{c}{$\langle$A,ns$\rangle$ $^2$S$^e$}& \multicolumn{4}{c}
{$\langle$A,np$\rangle$ $^2$P$^o$} \\ 
   & \multicolumn{2}{c}{$-$E(a.u.)} & \multicolumn{2}{c}{Excitation energy(eV)} 
   & \multicolumn{2}{c}{$-$E(a.u.)} & \multicolumn{2}{c}{Excitation energy(eV)} \\ 
\cline{2-3} \cline{4-5} \cline{6-7} \cline{8-9}
   & PW  & Ref. & PW & Ref.  
   & PW  & Ref. & PW & Ref.  \\
 2 &     &   &    &  
   & 2.2448  & 2.2503\footnotemark[1],2.247\footnotemark[2], & 142.385 
   & 142.255\footnotemark[1],142.439\footnotemark[3],142.12\footnotemark[4], \\
   &    &    &    &   
   &    & 2.2428\footnotemark[3] &   
   & 142.310\footnotemark[5],142.35\footnotemark[6],142.33\footnotemark[7]  \\
 3 & 1.9871  & 2.0048\footnotemark[3]$^,$\footnotemark[8], &  149.396 
   & 148.632\footnotemark[9],148.822\footnotemark[5],  
   & 1.9740  & 1.9935\footnotemark[10],1.991\footnotemark[2], & 149.753
   & 149.241\footnotemark[10],149.07\footnotemark[11],  \\
   &  & 2.0102\footnotemark[12] & & 148.788\footnotemark[12],148.914\footnotemark[3]
   &  & 1.9879\footnotemark[3]  & & 149.222\footnotemark[5],149.374\footnotemark[3] \\
 4 & 1.9402  & 1.9509\footnotemark[3], & 150.672  
   & 150.064\footnotemark[9],150.258\footnotemark[12],
   & 1.9321  & 1.9480\footnotemark[10],1.9423\footnotemark[3] & 150.893
   & 150.480\footnotemark[10],150.24\footnotemark[11], \\
   &   &  1.9561\footnotemark[12] &   & 
     150.381\footnotemark[3],150.181\footnotemark[5] &  &  &  
   & 150.615\footnotemark[3],150.397\footnotemark[5]  \\
 5 & 1.9194  & 1.9265\footnotemark[3] & 151.238
   & 150.600\footnotemark[9],151.045\footnotemark[3], 
   & 1.9160  & 1.9319\footnotemark[10],1.9241\footnotemark[3] & 151.331 
   & 150.917\footnotemark[10],150.67\footnotemark[11], \\
   &  &  &  & 150.765\footnotemark[5] &  &  &  
   & 151.110\footnotemark[3],150.836\footnotemark[5]  \\
 6 & 1.9094  & 1.9165\footnotemark[3] & 151.510  
   & 150.855\footnotemark[9],151.317\footnotemark[3],
   & 1.9075  & 1.9214\footnotemark[10],1.9145\footnotemark[3] & 151.562 
   & 151.203\footnotemark[10],150.88\footnotemark[11], \\
   &   &   &   &  151.025\footnotemark[5] &  &  &  
   & 151.371\footnotemark[3],151.057\footnotemark[5]  \\
 7 & 1.9037  & 1.9111\footnotemark[3] & 151.665
   & 151.001\footnotemark[9],151.464\footnotemark[3],
   & 1.9027  & 1.9160\footnotemark[10],1.9098\footnotemark[3]  & 151.692
   & 151.349\footnotemark[10],151.03\footnotemark[11] \\
   &  &  &  &  151.173\footnotemark[5] &  &  &  
   & 151.499\footnotemark[3],151.210\footnotemark[5] \\
 8 & 1.9004  & 1.9072\footnotemark[3] & 151.755  
   & 151.092\footnotemark[9],151.570\footnotemark[3],
   & 1.8996  & 1.9068\footnotemark[3] & 151.777 
   & 151.11\footnotemark[11],151.581\footnotemark[3], \\
   &  &   &  & 151.263\footnotemark[5]  &  &  &  & 151.285\footnotemark[5] \\
 9 & 1.8980  & 1.9052\footnotemark[3] & 151.820
   & 151.151\footnotemark[9],151.624\footnotemark[3],
   & 1.8976  & 1.9048\footnotemark[3] & 151.831 
   & 151.16\footnotemark[11],151.635\footnotemark[3], \\
   &  &  &  & 151.323\footnotemark[5] &  &  &  &  151.337\footnotemark[5] \\
10 & 1.8965  & 1.9037\footnotemark[3] & 151.861  
   & 151.190\footnotemark[9],151.665\footnotemark[3]
   & 1.8961  & 1.9034\footnotemark[3] & 151.872
   & 151.20\footnotemark[11],151.673\footnotemark[3]  \\
11 & 1.8953  & 1.9026\footnotemark[3] & 151.894
   & 151.226\footnotemark[9],151.695\footnotemark[3]
   & 1.8950  & 1.9024\footnotemark[3] & 151.902 & 151.701\footnotemark[3]  \\
12 & 1.8944  & 1.9018\footnotemark[3] & 151.918  
   & 151.247\footnotemark[9],151.717\footnotemark[3]
   & 1.8943  &    & 151.921 &     \\
13 & 1.8938  &    & 151.935 & 151.264\footnotemark[9]
   & 1.8937  &    & 151.937 &     \\
14 & 1.8933  &    & 151.948 & 151.274\footnotemark[9]
   & 1.8932  &    & 151.951 &     \\
15 & 1.8929  &    & 151.959 & 151.288\footnotemark[9]
   & 1.8928  &    & 151.962 &     \\
16 & 1.8925  &    & 151.970 & 151.296\footnotemark[9]
   & 1.8925  &    & 151.970 &     \\
17 & 1.8923  &    & 151.975 & 151.303\footnotemark[9] 
   & 1.8923  &    & 151.975 &     \\
18 & 1.8921  &    & 151.981 & 151.309\footnotemark[9] 
   & 1.8921  &    & 151.981 &     \\
19 & 1.8919  &    & 151.986 & 151.314\footnotemark[9] 
   & 1.8919  &    & 151.986 &     \\     
20 & 1.8917  &    & 151.992 & 151.318\footnotemark[9]  
   & 1.8918  &    & 151.989 &     \\     
21 & 1.8915  &    & 151.997 & 151.322\footnotemark[9]    
   & 1.8917  &    & 151.992 &     \\     
22 & 1.8914  &    & 152.000 & 151.325\footnotemark[9]
   & 1.8916  &    & 151.994 &      \\     
\end{tabular}
\end{ruledtabular}
\footnotetext[1]{Reference [23].} 
\footnotetext[2]{Reference [53].} 
\footnotetext[3]{Reference [33].} 
\footnotetext[4]{Reference [10].} 
\footnotetext[5]{Reference [28].}
\footnotetext[6]{Reference [6].}
\footnotetext[7]{Reference [7].}
\footnotetext[8]{Reference [35].}
\footnotetext[9]{Reference [31].}
\footnotetext[10]{Reference [24].}
\footnotetext[11]{Reference [29].}
\footnotetext[12]{Reference [25].}
\end{table}
\endgroup

Now in Table II, we compare the even-parity $\langle$A,nd$\rangle$ and 
$\langle$D,ns$\rangle$ $^2$D$^e$ hollow states of Li arising from electronic 
configurations 2s$^2$nd and 2p$^2$ns having n=24 and 25 respectively with the existing
literature data. No experimental results have been reported so far for the former 
series to the best of our knowledge, whereas only the lowest state of the latter 
series has been observed. It lies at 144.77 eV in the photoelectron spectroscopy [11]
and our result matches excellently with this (the excitation energy is only 0.043 eV
lower than the experimental values with a deviation of 0.03\%). Both the 
hyperspherical coordinate approach [35] as well as the TDM [33] give the energy of 
2s$^2$3d state at $-$1.9614 a.u., whereas the CCR value is $-$1.9659 a.u. [25]. These
are to be compared with the present value of $-$1.9461 a.u. The CCR results are also
available for the $\langle$D,ns$\rangle$ series having n=2,3. It is noticed that 
for the former series, the state energies are underestimated in all cases with respect
to the TDM results, whereas for the latter, the same is overestimated for n=2,4--7
(n=8 shows complete agreement). This overestimation could occur either because of (a)
the nonvariational nature of the exchange potential employed and/or (b) the inadequacy
of the LYP correlation energy functional used. Resonances up to 9 for both the series 
were earlier reported in the R-matrix calculation [28]. The absolute per cent deviation
of the $\langle$A,nd$\rangle$ (n=3--11) and $\langle$D,ns$\rangle$ (n=2--9) state
energies are 0.44--0.78\% and 0.00--2.15\% respectively compared to the TDM values 
[33]. We note that $\langle$D,ns$\rangle$ gives the largest deviation in the energy in
our calculation, (2.15\% for n=3) as well as the smallest deviation (0.00\% for n=8). 
However it is also noted that significantly varied and contrasting excitation energies 
were reported in the literature for this resonance state, thus requiring more 
extensive and elaborate computations to determine its position with greater confidence.
Recently accurate excitation energies corresponding to the higher resonances of both 
the series (n up to 22 and 25 for the former and latter series respectively) have been
reported in the eigenphase derivative theory coupled with the R-matrix method, and the
present results show absolute discrepancies in the range of 0.46--0.50\% and 
0.04--0.58\% for the two series. 

\begingroup
\squeezetable
\begin{table}
\caption {\label{tab:table2}Comparison of the state (in a.u.) and excitation energies
(in eV) relative to the Li ground state [54], of 2s$^2$nd and 2p$^2$ns $^2$D$^e$ 
resonances of Li. PW signifies present work.}
\begin{ruledtabular}
\begin{tabular}{lllllllll}
 n  & \multicolumn{4}{c}{$\langle$A,nd$\rangle$ $^2$D$^e$}& 
      \multicolumn{4}{c}{$\langle$D,ns$\rangle$ $^2$D$^e$} \\ 
   & \multicolumn{2}{c}{$-$E(a.u.)} & \multicolumn{2}{c}{Excitation energy(eV)} 
   & \multicolumn{2}{c}{$-$E(a.u.)} & \multicolumn{2}{c}{Excitation energy(eV)} \\ 
\cline{2-3} \cline{4-5} \cline{6-7} \cline{8-9}
   & PW  & Ref. & PW & Ref.  
   & PW  & Ref. & PW & Ref.  \\
2  &    &     &     &      
   & 2.1587   & 2.1582\footnotemark[5], & 144.727
   & 144.762\footnotemark[5],144.82\footnotemark[6],144.77\footnotemark[7],  \\
   &    &     &     &     &    &   2.1480\footnotemark[2]
   &  & 145.018\footnotemark[2],144.664\footnotemark[4]        \\
3  & 1.9461   & 1.9614\footnotemark[1]$^,$\footnotemark[2],
   & 150.512  & 149.826\footnotemark[4],150.095\footnotemark[2],
   & 1.8458   & 1.8489\footnotemark[3], & 153.241 
   & 153.180\footnotemark[3],152.543\footnotemark[6],                      \\
   &    & 1.9659\footnotemark[3] &  & 149.982\footnotemark[6] &     
   & 1.8069\footnotemark[2] & & 152.364\footnotemark[4],154.299\footnotemark[2] \\
4  & 1.9202   & 1.9319\footnotemark[2] &  151.216 
   & 150.484\footnotemark[4],150.898\footnotemark[2],
   & 1.7955   & 1.7782\footnotemark[2] &  154.609  
   & 153.929\footnotemark[4],155.080\footnotemark[2], \\
   &  &  &  & 150.651\footnotemark[6] &  &  &  & 154.050\footnotemark[6] \\
5  & 1.9100   & 1.9193\footnotemark[2] &  151.494
   & 150.798\footnotemark[4],151.241\footnotemark[2],
   & 1.7765   & 1.7694\footnotemark[2] &  155.126
   & 154.517\footnotemark[4],155.319\footnotemark[2], \\  
   &  &  &  & 150.965\footnotemark[6] &  &  &  & 154.673\footnotemark[6] \\
6  & 1.9035   & 1.9120\footnotemark[2] &  151.671
   & 150.968\footnotemark[4],151.439\footnotemark[2],
   & 1.7672   & 1.7645\footnotemark[2] &  155.379
   & 154.741\footnotemark[4],155.453\footnotemark[2], \\
   &  &  &  & 151.137\footnotemark[6] &  &  &  & 154.881\footnotemark[6] \\
7  & 1.8995   & 1.9084\footnotemark[2] &  151.780 
   & 151.069\footnotemark[4],151.537\footnotemark[2],
   & 1.7618   & 1.7607\footnotemark[2] &  155.526 
   & 154.895\footnotemark[4],155.556\footnotemark[2], \\
   &  &  &  & 151.239\footnotemark[6] &  &  &  & 155.013\footnotemark[6] \\
8  & 1.8969   & 1.9059\footnotemark[2] &  151.850
   & 151.141\footnotemark[4],151.605\footnotemark[2],
   & 1.7586   & 1.7586\footnotemark[2] &  155.613
   & 154.971\footnotemark[4],155.613\footnotemark[2], \\
   &  &  &  & 151.312\footnotemark[6] &  &  &  & 155.118\footnotemark[6] \\
9  & 1.8951   & 1.9042\footnotemark[2] &  151.899 
   & 151.185\footnotemark[4],151.652\footnotemark[2], 
   & 1.7564   & 1.7571\footnotemark[2] &  155.673 
   & 155.026\footnotemark[4],155.654\footnotemark[2], \\
   &  &  &  & 151.356\footnotemark[6] &  &  &  & 155.172\footnotemark[6] \\
10 & 1.8936   & 1.9029\footnotemark[2] &  151.940
   & 151.217\footnotemark[4],151.687\footnotemark[2] 
   & 1.7549   &       & 155.714    & 155.065\footnotemark[4]     \\
11 & 1.8924   & 1.9020\footnotemark[2] &  151.973 
   & 151.241\footnotemark[4],151.712\footnotemark[2]
   & 1.7538   &       & 155.744    & 155.102\footnotemark[4]     \\
12 & 1.8914   &       & 152.000    & 151.260\footnotemark[4]
   & 1.7530   &       & 155.765    & 155.122\footnotemark[4]     \\
13 & 1.8906   &       & 152.022    & 151.274\footnotemark[4] 
   & 1.7524   &       & 155.782    & 155.138\footnotemark[4]     \\
14 & 1.8901   &       & 152.035    & 151.285\footnotemark[4]
   & 1.7519   &       & 155.795    & 155.151\footnotemark[4]     \\
15 & 1.8897   &       & 152.046    & 151.294\footnotemark[4]
   & 1.7515   &       & 155.806    & 155.161\footnotemark[4]     \\
16 & 1.8893   &       & 152.057    & 151.302\footnotemark[4]
   & 1.7512   &       & 155.814    & 155.169\footnotemark[4]     \\
17 & 1.8890   &       & 152.065    & 151.308\footnotemark[4]
   & 1.7509   &       & 155.823    & 155.176\footnotemark[4]     \\
18 & 1.8888   &       & 152.071    & 151.313\footnotemark[4] 
   & 1.7507   &       & 155.828    & 155.182\footnotemark[4]     \\
19 & 1.8886   &       & 152.076    & 151.317\footnotemark[4]
   & 1.7505   &       & 155.833    & 155.187\footnotemark[4]     \\
20 & 1.8884   &       & 152.082    & 151.321\footnotemark[4]
   & 1.7504   &       & 155.836    & 155.191\footnotemark[4]     \\
21 & 1.8883   &       & 152.084    & 151.324\footnotemark[4]
   & 1.7502   &       & 155.842    & 155.194\footnotemark[4]     \\
22 & 1.8882   &       & 152.087    & 151.327\footnotemark[4]
   & 1.7501   &       & 155.844    & 155.197\footnotemark[4]     \\
23 & 1.8881   &       & 152.090    & 
   & 1.7500   &       & 155.847    & 155.200\footnotemark[4]     \\
24 & 1.8880   &       & 152.094    &
   & 1.7499   &       & 155.850    & 155.202\footnotemark[4]     \\
25 &          &       &            &
   & 1.7498   &       & 155.853    & 155.205\footnotemark[4]     \\
\end{tabular}
\end{ruledtabular}
\footnotetext[1]{Reference [35].}
\footnotetext[2]{Reference [33].} 
\footnotetext[3]{Reference [25].}
\footnotetext[4]{Reference [31].}
\footnotetext[5]{Reference [23].} 
\footnotetext[6]{Reference [28].}
\footnotetext[7]{Reference [11].}
\end{table}
\endgroup

Next in table III we turn into a comparison of the calculated density functional 
results for the even- and odd-parity quartet P states arising from the 2p$^2$ns 
$\langle$C,ns $\rangle$, n=2--24 and 2s2pns $\langle$B,ns$\rangle$, n=3--22 with the
available literature data. Experimental results are yet to be obtained for any of 
these states, while from a theoretical viewpoint, these are relatively less explored 
compared to the previously discussed hollow states. The energy of the lowest state of
the even series in our calculation is only 0.004 a.u., above the CCR result [23], 
while the same for the odd series is only 0.0037 a.u., below the hyperspherical 
coordinate result [35] indicating the accuracy in the present method. Both these 
resonances have been reported through R-matrix calculation for n up to 9 [28] as well 
as in the TDM calculation for n up to 12 [33] and the current results show excellent 
agreement with both of these. The $\langle$B,ns$\rangle$ series show overestimations in
energies for all n compared to the TDM result [33], but the $\langle$C,ns $\rangle$ 
series show underestimations as well for some of the states. For the moderately 
high-lying $\langle$C,ns$\rangle$ states with n$\ge$6, we notice almost complete 
agreement in the energies in our calculation with the TDM values [33]. The calculated 
energies for the even- and odd-parity states are 0.00--0.27\% and 0.19--0.65\% above 
and below the TDM values for up to the 12th resonance respectively. The deviations in 
the excitation energies are 0.03--0.17\% and 0.02-0.10\% for the two series relative 
to the R-matrix results [28]. Resonances above n=12 are reported here for the first 
time for both these series. 

\begingroup
\squeezetable
\begin{table}
\caption {\label{tab:table3}Comparison of the state (in a.u.) and excitation energies
(in eV) relative to the Li ground state [54], of 2p$^2$ns $^4$P$^e$ and 2s2pns 
$^4$P$^o$ resonances of Li. PW signifies present work.}
\begin{ruledtabular}
\begin{tabular}{lllllllll}
 n  & \multicolumn{4}{c}{$\langle$C,ns$\rangle$ $^4$P$^e$} & 
      \multicolumn{4}{c}{$\langle$B,ns$\rangle$ $^4$P$^o$} \\ 
   & \multicolumn{2}{c}{$-$E(a.u.)} & \multicolumn{2}{c}{Excitation energy(eV)} 
   & \multicolumn{2}{c}{$-$E(a.u.)} & \multicolumn{2}{c}{Excitation energy(eV)} \\ 
\cline{2-3} \cline{4-5} \cline{6-7} \cline{8-9}
   & PW  & Ref. & PW & Ref.  
   & PW  & Ref. & PW & Ref.  \\
2  & 2.2390 & 2.2394\footnotemark[1],2.2331\footnotemark[2] & 142.543 
   & 142.553\footnotemark[1],142.59\footnotemark[3],
   &         &        &        &        \\
   &    &    &   & 142.703\footnotemark[2]   &    &    &     &     \\   
3  & 1.8838  & 1.8889\footnotemark[4]$^,$\footnotemark[2]    & 152.207
   & 152.068\footnotemark[2],151.976\footnotemark[3] 
   & 1.9878  & 1.9841\footnotemark[4]$^,$\footnotemark[2]
   & 149.377 & 149.478\footnotemark[2],149.409\footnotemark[3]      \\
4  & 1.8319  & 1.8353\footnotemark[2]  & 153.619
   & 153.526\footnotemark[2],153.365\footnotemark[3]
   & 1.9312  & 1.9228\footnotemark[2]  & 150.917 
   & 151.146\footnotemark[2],151.045\footnotemark[3]  \\
5  & 1.8127  & 1.8131\footnotemark[2]  & 154.141 
   & 154.130\footnotemark[2],153.892\footnotemark[3]
   & 1.9108  & 1.8992\footnotemark[2]  & 151.472 
   & 151.788\footnotemark[2],151.619\footnotemark[3]  \\
6  & 1.8033  & 1.8034\footnotemark[2]  & 154.397 
   & 154.394\footnotemark[2],154.147\footnotemark[3]
   & 1.9011  & 1.8890\footnotemark[2]  & 151.736 
   & 152.065\footnotemark[2],151.889\footnotemark[3]  \\
7  & 1.7979  & 1.7980\footnotemark[2]  & 154.544 
   & 154.541\footnotemark[2],154.290\footnotemark[3]
   & 1.8956  & 1.8834\footnotemark[2]  & 151.886 
   & 152.218\footnotemark[2],152.039\footnotemark[3]  \\
8  & 1.7947  & 1.7947\footnotemark[2]  & 154.631 
   & 154.631\footnotemark[2],154.378\footnotemark[3]
   & 1.8922  & 1.8800\footnotemark[2]  & 151.978 
   & 152.310\footnotemark[2],152.131\footnotemark[3]  \\
9  & 1.7925  & 1.7927\footnotemark[2]  & 154.691 
   & 154.685\footnotemark[2],154.436\footnotemark[3]
   & 1.8899  & 1.8778\footnotemark[2]  & 152.041 
   & 152.370\footnotemark[2],152.191\footnotemark[3]  \\
10 & 1.7909  & 1.7909\footnotemark[2]  & 154.734 & 154.734\footnotemark[2] 
   & 1.8884  & 1.8762\footnotemark[2]  & 152.082 & 152.413\footnotemark[2]  \\
11 & 1.7898  & 1.7899\footnotemark[2]  & 154.764 & 154.761\footnotemark[2] 
   & 1.8873  & 1.8751\footnotemark[2]  & 152.111 & 152.443\footnotemark[2]  \\
12 & 1.7890  & 1.7891\footnotemark[2]  & 154.786 & 154.783\footnotemark[2] 
   & 1.8864  & 1.8743\footnotemark[2]  & 152.136 & 152.465\footnotemark[2]  \\
13 & 1.7884  &                         & 154.802 &                         
   & 1.8858  &                         & 152.152 &                          \\
14 & 1.7879  &                         & 154.816 &                         
   & 1.8853  &                         & 152.166 &                          \\
15 & 1.7875  &                         & 154.827 &                         
   & 1.8849  &                         & 152.177 &                          \\
16 & 1.7872  &                         & 154.835 &                         
   & 1.8846  &                         & 152.185 &                          \\
17 & 1.7869  &                         & 154.843 &                         
   & 1.8843  &                         & 152.193 &                          \\
18 & 1.7867  &                         & 154.849 &                         
   & 1.8841  &                         & 152.199 &                          \\
19 & 1.7865  &                         & 154.854 &                         
   & 1.8839  &                         & 152.204 &                          \\
20 & 1.7864  &                         & 154.857 &                         
   & 1.8838  &                         & 152.207 &                          \\
21 & 1.7862  &                         & 154.862 &                         
   & 1.8836  &                         & 152.212 &                          \\
22 & 1.7861  &                         & 154.865 &                         
   & 1.8835  &                         & 152.215 &                          \\
23 & 1.7860  &                         & 154.868 &                         
   &         &                         &         &                          \\
24 & 1.7859  &                         & 154.870 &                         
   &         &                         &         &                          \\
\end{tabular}
\end{ruledtabular}
\footnotetext[1]{Reference [23].} 
\footnotetext[2]{Reference [33].} 
\footnotetext[3]{Reference [28].}
\footnotetext[4]{Reference [35].}
\end{table}
\endgroup

The even- and odd-parity 2s2pnp $\langle$B,np$\rangle$, n=3--22 and 2p$^2$np 
$\langle$C,np$\rangle$, n=3--20 $^4$D hollow resonances of Li are calculated and 
compared in Table IV. No experimental results have been reported as yet for any of 
these states, and the theoretical results are quite scanty. The calculated lowest 
state energies of the two series are 0.23\% and 0.02\% below both the hyperspherical 
[35] as well as TDM results [33] respectively. The density functional state energies 
for the even-parity resonances (up to n$\le$12) match well with the TDM values with
overestimations in the range of 0.23--0.65\%, whereas the same of the odd-parity states
match excellently (both overestimations and underestimations are noticed) with those
having deviations from 0.006--0.02\%. The discrepancies in the calculated 
excitation energies remain within 0.04--0.10\% and 0.11--0.17\% for the even- and 
odd-parity states with respect to the R-matrix values [28]. The excitation energies 
up to the 12th resonance for both the series, calculated using the TDM calculation 
[33] are also quoted in the table and the present results show satisfactory 
agreement. No reference results are found in the literature for the resonances with 
n$\ge$13. 

\begingroup
\squeezetable
\begin{table}
\caption {\label{tab:table4}Comparison of the state (in a.u.) and excitation energies
(in eV) relative to the Li ground state [54], of 2s2pnp $^4$D$^e$ and 2p$^2$np 
$^4$D$^o$ resonances of Li. PW signifies present work.}
\begin{ruledtabular}
\begin{tabular}{lllllllll}
 n  & \multicolumn{4}{c}{$\langle$B,np$\rangle$ $^4$D$^e$}& 
      \multicolumn{4}{c}{$\langle$C,np$\rangle$ $^4$D$^o$} \\ 
   & \multicolumn{2}{c}{$-$E(a.u.)} & \multicolumn{2}{c}{Excitation energy(eV)} 
   & \multicolumn{2}{c}{$-$E(a.u.)} & \multicolumn{2}{c}{Excitation energy(eV)} \\ 
\cline{2-3} \cline{4-5} \cline{6-7} \cline{8-9}
   & PW  & Ref. & PW & Ref.  
   & PW  & Ref. & PW & Ref.  \\
3  & 1.9705    & 1.9659\footnotemark[1]$^,$\footnotemark[2] 
   & 149.848   & 149.973\footnotemark[2],149.907\footnotemark[3] 
   & 1.8727    & 1.8723\footnotemark[1]$^,$\footnotemark[2]
   & 152.509   & 152.520\footnotemark[2],152.337\footnotemark[3]   \\
4  & 1.9251    & 1.9168\footnotemark[2] & 151.083  
   & 151.309\footnotemark[2],151.197\footnotemark[3]
   & 1.8276    & 1.8292\footnotemark[2] & 153.736  
   & 153.692\footnotemark[2],153.500\footnotemark[3]  \\
5  & 1.9081    & 1.8969\footnotemark[2] & 151.546  
   & 151.850\footnotemark[2],151.685\footnotemark[3]
   & 1.8106    & 1.8111\footnotemark[2] & 154.198  
   & 154.185\footnotemark[2],153.952\footnotemark[3]  \\
6  & 1.8996    & 1.8877\footnotemark[2] & 151.777  
   & 152.101\footnotemark[2],151.924\footnotemark[3]
   & 1.8021    & 1.8023\footnotemark[2] & 154.430  
   & 154.424\footnotemark[2],154.179\footnotemark[3]  \\
7  & 1.8947    & 1.8826\footnotemark[2] & 151.910  
   & 152.239\footnotemark[2],152.060\footnotemark[3]
   & 1.7973    & 1.7974\footnotemark[2] & 154.557  
   & 154.557\footnotemark[2],154.309\footnotemark[3]  \\
8  & 1.8916    & 1.8795\footnotemark[2] & 151.994  
   & 152.324\footnotemark[2],152.145\footnotemark[3]
   & 1.7942    & 1.7943\footnotemark[2] & 154.644  
   & 154.642\footnotemark[2],154.390\footnotemark[3]  \\
9  & 1.8896    & 1.8774\footnotemark[2] & 152.049  
   & 152.381\footnotemark[2],152.201\footnotemark[3]
   & 1.7922    & 1.7923\footnotemark[2] & 154.699  
   & 154.696\footnotemark[2],154.444\footnotemark[3]  \\
10 & 1.8881    & 1.8760\footnotemark[2] & 152.090  & 152.419\footnotemark[2]
   & 1.7907    & 1.7908\footnotemark[2] & 154.740  & 154.737\footnotemark[2]  \\
11 & 1.8871    & 1.8749\footnotemark[2] & 152.117  & 152.449\footnotemark[2]
   & 1.7897    & 1.7898\footnotemark[2] & 154.767  & 154.764\footnotemark[2]  \\
12 & 1.8863    & 1.8741\footnotemark[2] & 152.139  & 152.471\footnotemark[2]
   & 1.7889    & 1.7890\footnotemark[2] & 154.789  & 154.790\footnotemark[2]  \\
13 & 1.8857    &      & 152.155  &      & 1.7883  &       & 154.805  &   \\
14 & 1.8852    &      & 152.169  &      & 1.7878  &       & 154.819  &   \\
15 & 1.8848    &      & 152.179  &      & 1.7874  &       & 154.829  &   \\
16 & 1.8845    &      & 152.188  &      & 1.7871  &       & 154.838  &   \\
17 & 1.8843    &      & 152.193  &      & 1.7869  &       & 154.843  &   \\
18 & 1.8841    &      & 152.199  &      & 1.7867  &       & 154.849  &   \\
19 & 1.8840    &      & 152.201  &      & 1.7866  &       & 154.851  &   \\
20 & 1.8839    &      & 152.204  &      & 1.7865  &       & 154.854  &   \\
21 & 1.8838    &      & 152.207  &      &         &       &          &   \\
22 & 1.8837    &      & 152.209  &      &         &       &          &   \\
\end{tabular}
\end{ruledtabular}
\footnotetext[1]{Reference [35].}
\footnotetext[2]{Reference [33].} 
\footnotetext[3]{Reference [28].}
\end{table}
\endgroup

Table V now compares the computed state- and excitation energies for the 2s2pnd
$\langle$C,nd$\rangle$ $^4$F$^o$, n=3--20 and 2p$^2$nd $\langle$B,nd$\rangle$ 
$^4$F$^e$, n=3--24 hollow resonances of Li with the existing literature data. Once  
again, no experimental results are yet available for these resonances 
and very few theoretical results have been reported so far. The calculated lowest state
energies of the two series show good agreement with the hyperspherical [35] as well as 
the TDM [33] energies having absolute deviations 0.19\% and 0.23\% respectively;  
the former is overestimated, while the latter underestimated. The overall agreements in
the state energies with respect to the TDM results for the former and latter resonances
are good (overestimations by 0.19--0.46\%) and excellent (underestimations by 
0.02--0.23\%) respectively for n up to 11 and 12 respectively. The excitation energies
match quite decently with the R-matrix results [28] having absolute deviations in the 
ranges 0.03--0.04\% and 0.16--0.19\% (for resonances up to 9) as well as with the TDM 
values [33]. Resonances with n$>$12 for both of these series are calculated here for 
the first time. 

\begingroup
\squeezetable
\begin{table}
\caption {\label{tab:table5}Comparison of the state (in a.u.) and excitation energies
(in eV) relative to the Li ground state [54], of 2s2pnd $^4$F$^o$ and 2p$^2$nd 
$^4$F$^e$ resonances of Li. PW signifies present work.}
\begin{ruledtabular}
\begin{tabular}{lllllllll}
 n  & \multicolumn{4}{c}{$\langle$C,nd$\rangle$ $^4$F$^o$} & 
      \multicolumn{4}{c}{$\langle$B,nd$\rangle$ $^4$F$^e$} \\ 
   & \multicolumn{2}{c}{$-$E(a.u.)} & \multicolumn{2}{c}{Excitation energy(eV)} 
   & \multicolumn{2}{c}{$-$E(a.u.)} & \multicolumn{2}{c}{Excitation energy(eV)} \\ 
\cline{2-3} \cline{4-5} \cline{6-7} \cline{8-9}
   & PW  & Ref. & PW & Ref.  
   & PW  & Ref. & PW & Ref.  \\
3  & 1.9397  & 1.9360\footnotemark[1]$^,$\footnotemark[2] 
   & 150.686 & 150.786\footnotemark[2],150.723\footnotemark[3]
   & 1.8456  & 1.8498\footnotemark[1]$^,$\footnotemark[2]
   & 153.246 & 153.132\footnotemark[2],153.004\footnotemark[3]      \\
4  & 1.9128  & 1.9061\footnotemark[2] & 151.418 
   & 151.600\footnotemark[2],151.465\footnotemark[3]
   & 1.8181  & 1.8215\footnotemark[2] & 153.994 
   & 153.902\footnotemark[2],153.705\footnotemark[3]   \\
5  & 1.9010  & 1.8925\footnotemark[2] & 151.739 
   & 151.970\footnotemark[2],151.807\footnotemark[3]
   & 1.8070  & 1.8081\footnotemark[2] & 154.296 
   & 154.266\footnotemark[2],154.045\footnotemark[3]   \\
6  & 1.8940  & 1.8854\footnotemark[2] & 151.929 
   & 152.163\footnotemark[2],151.991\footnotemark[3]
   & 1.8002  & 1.8009\footnotemark[2] & 154.481 
   & 154.462\footnotemark[2],154.229\footnotemark[3]   \\
7  & 1.8898  & 1.8812\footnotemark[2] & 152.043 
   & 152.277\footnotemark[2],152.101\footnotemark[3]
   & 1.7961  & 1.7966\footnotemark[2] & 154.593 
   & 154.579\footnotemark[2],154.339\footnotemark[3]   \\
8  & 1.8871  & 1.8786\footnotemark[2] & 152.117 
   & 152.348\footnotemark[2],152.171\footnotemark[3]
   & 1.7935  & 1.7939\footnotemark[2] & 154.664 
   & 154.653\footnotemark[2],154.410\footnotemark[3]   \\
9  & 1.8853  & 1.8768\footnotemark[2] & 152.166 
   & 152.397\footnotemark[2],152.210\footnotemark[3]
   & 1.7916  & 1.7921\footnotemark[2] & 154.715 
   & 154.702\footnotemark[2],154.458\footnotemark[3]   \\
10 & 1.8840  & 1.8755\footnotemark[2] & 152.201 & 152.433\footnotemark[2]
   & 1.7904  & 1.7907\footnotemark[2] & 154.748 & 154.740\footnotemark[2]   \\
11 & 1.8830  & 1.8746\footnotemark[2] & 152.228 & 152.457\footnotemark[2]
   & 1.7894  & 1.7898\footnotemark[2] & 154.764 & 154.764\footnotemark[2]   \\
12 & 1.8822  &                        & 152.250 &                        
   & 1.7887  & 1.7890\footnotemark[2] & 154.794 & 154.786\footnotemark[2]   \\
13 & 1.8816  &    & 152.267 &         & 1.7881  &  & 154.810 &    \\
14 & 1.8811  &    & 152.280 &         & 1.7877  &  & 154.821 &    \\
15 & 1.8807  &    & 152.291 &         & 1.7873  &  & 154.832 &    \\
16 & 1.8804  &    & 152.299 &         & 1.7871  &  & 154.838 &    \\
17 & 1.8802  &    & 152.305 &         & 1.7868  &  & 154.846 &    \\
18 & 1.8800  &    & 152.310 &         & 1.7866  &  & 154.851 &    \\
19 & 1.8799  &    & 152.313 &         & 1.7864  &  & 154.857 &    \\
20 & 1.8798  &    & 152.316 &         & 1.7863  &  & 154.859 &    \\
21 &         &    &         &         & 1.7862  &  & 154.862 &    \\
22 &         &    &         &         & 1.7861  &  & 154.865 &    \\
23 &         &    &         &         & 1.7860  &  & 154.868 &    \\
24 &         &    &         &         & 1.7859  &  & 154.870 &    \\

\end{tabular}
\end{ruledtabular}
\footnotetext[1]{Reference [35].}
\footnotetext[2]{Reference [33].} 
\footnotetext[3]{Reference [28].}
\end{table}
\endgroup

Finally table VI computes and compares the last of the $2l2l'$n$l''$ hollow resonance
series of Li considered in this work, {\em viz.}, 2p$^2$np $\langle$D,np$\rangle$ 
$^2$F$^o$, n=3--23 and 2p$^2$nd $\langle$D,nd$\rangle$ $^2$G$^e$, n=3--22 respectively.
Once again experimental results are as yet unavailable for both of these series. No 
theoretical results are reported for the latter series, whereas the same for the former
series is quite scarce. The present computed term energy of the lowest $^2$F$^o$ 
resonance is in good agreement with the hyperspherical [35] as well as the TDM [33] 
calculations (underestimated by 0.42\%). The state energies remain above by 
0.14--0.46\% from the TDM values [33] and the absolute deviations in excitation 
energies are 0.28--0.35\% with respect to the R-matrix results [28]. 

\begingroup
\squeezetable
\begin{table}
\caption {\label{tab:table6}Comparison of the state (in a.u.) and excitation energies
(in eV) relative to the Li ground state [54], of 2p$^2$np $^2$F$^o$ and 2p$^2$nd 
$^2$G$^e$ resonances of Li. PW signifies present work.}
\begin{ruledtabular}
\begin{tabular}{lllllllll}
 n  & \multicolumn{4}{c}{$\langle$D,np$\rangle$ $^2$F$^o$} & 
      \multicolumn{4}{c}{$\langle$D,nd$\rangle$ $^2$G$^e$} \\ 
   & \multicolumn{2}{c}{$-$E(a.u.)} & \multicolumn{2}{c}{Excitation energy(eV)} 
   & \multicolumn{2}{c}{$-$E(a.u.)} & \multicolumn{2}{c}{Excitation energy(eV)} \\ 
\cline{2-3} \cline{4-5} \cline{6-7} \cline{8-9}
   & PW  & Ref. & PW & Ref.  
   & PW  & Ref. & PW & Ref.  \\
3  & 1.8429  & 1.8506\footnotemark[1]$^,$\footnotemark[2]  
   & 153.319 & 153.110\footnotemark[2],152.892\footnotemark[3] 
   & 1.8153  &     &  154.070   &        \\
4  & 1.7908  & 1.7991\footnotemark[2]  & 154.737 
   & 154.511\footnotemark[2],154.240\footnotemark[3]
   & 1.7808  &     &  155.009   &        \\
5  & 1.7743  & 1.7787\footnotemark[2]  & 155.186 
   & 155.066\footnotemark[2],154.651\footnotemark[3]
   & 1.7703  &     &  155.295   &        \\
6  & 1.7660  & 1.7686\footnotemark[2]  & 155.412 
   & 155.341\footnotemark[2],154.921\footnotemark[3]
   & 1.7639  &     &  155.469   &        \\
7  & 1.7611  & 1.7636\footnotemark[2]  & 155.545 
   & 155.477\footnotemark[2],155.039\footnotemark[3]
   & 1.7598  &     &  155.580   &        \\
8  & 1.7581  & 1.7608\footnotemark[2]  & 155.627 
   & 155.553\footnotemark[2],155.117\footnotemark[3]
   & 1.7573  &     &  155.648   &        \\
9  & 1.7561  &                         & 155.681 
   & 155.185\footnotemark[3]                         
   & 1.7555  &     &  155.697   &        \\
10 & 1.7545  & 1.7583\footnotemark[2]  & 155.725 & 155.621\footnotemark[2] 
   & 1.7543  &     &  155.730   &        \\
11 & 1.7536  & 1.7569\footnotemark[2]  & 155.749 & 155.659\footnotemark[2] 
   & 1.7534  &     &  155.755   &        \\
12 & 1.7529  &     & 155.768    &      & 1.7527  &     &  155.774   &     \\
13 & 1.7523  &     & 155.784    &      & 1.7521  &     &  155.790   &     \\
14 & 1.7518  &     & 155.798    &      & 1.7516  &     &  155.804   &     \\
15 & 1.7514  &     & 155.809    &      & 1.7512  &     &  155.814   &     \\
16 & 1.7511  &     & 155.817    &      & 1.7509  &     &  155.823   &     \\
17 & 1.7509  &     & 155.823    &      & 1.7507  &     &  155.828   &     \\
18 & 1.7507  &     & 155.828    &      & 1.7505  &     &  155.833   &     \\
19 & 1.7505  &     & 155.833    &      & 1.7503  &     &  155.839   &     \\
20 & 1.7503  &     & 155.839    &      & 1.7501  &     &  155.844   &     \\
21 & 1.7502  &     & 155.842    &      & 1.7500  &     &  155.844   &     \\
22 & 1.7501  &     & 155.844    &      & 1.7499  &     &  155.850   &     \\
23 & 1.7500  &     & 155.847    &      &         &     &            &     \\
\end{tabular}
\end{ruledtabular}
\footnotetext[1]{Reference [35].}
\footnotetext[2]{Reference [33].} 
\footnotetext[3]{Reference [28].}
\end{table}
\endgroup

As a further extension of the present method, table VII gives results for several
higher lying triply excited hollow resonances of Li in which both the K and L shells
are empty, the so-called doubly hollow states, {\em viz.,} $3l3l'$n$l''$(3$\le$n$\le$6)
($^2$S$^e$, $^2$P$^o$, $^2$D$^e$, $^2$F$^o$, $^4$S$^o$, $^4$P$^{e,o}$, $^4$D$^{e,o}$
and $^4$F$^o$). Although a decent number of accurate, reliable experimental and 
theoretical results are available for the $2l2l'$n$l''$ resonances of Li as already 
discussed in tables I through VI, the same for resonances having all the three 
electrons residing in shells with principal quantum numbers 3 or more, are very limited
presumably because of the greater challenges encountered. Some distinctive features of 
these resonances are: they are weak (by about an order of magnitude compared to the 
previously discussed hollow states), broad and having much larger widths [8]. The 
principal difficulties in dealing with such higher hollow states at larger photon 
energies are mainly due to a very rapid increase in the density of possible triply 
excited states and the lower states of same symmetry, as well as of the number of open 
channels available, leading to very strong and quite complicated electron correlation 
effects. In the CCR calculation, this might require cumbersome construction and 
diagonalization of complex matrices of very large order to search for the behavior of 
a great number of roots in the complex energy plane. On the other hand, in a FD type 
numerical calculation, this often leads to the usual convergence problems. Despite all 
these, some attempts have been made in the recent past to measure and calculate these 
states. For example, the energies and decay rates of N$^{4+}$ and N$^{2+}$$3l3l'$3$l''$
configurations using a CI approach [22]; positions and widths of N$^{4+}$ (3,3,3)
$^2$S$^{e,o}$ states employing the space partition as well as a stabilization procedure
both of which use the L$^2$  discretization [26]; large scale state specific theory 
calculations for 11 n=3 resonances of He$^-$ accounting for all localized electron 
correlations [21]; critical issues in the theory and computation of the lowest three 
n=3 intrashell states, {\em viz}., 3s$^2$3p $^2$P$^o$, 3s3p$^2$ $^4$P$^e$ and 3s3p$^2$ 
$^2$D$^e$ of Z=2--7 in the light of state specific theory [55]; energies, widths and 
Auger branching ratios for eight He$^-$ $3l3l'$3$l''$ states using the CCR method [55], 
etc. A semiquantitative analysis of the angular correlation of 64 n=3 intrashell 
states of a model three-electron atom confined on the surface of a sphere were 
presented recently [57]. The only triply photoexcited (3,3,3) KL-hollow states studied
so far for Li are the 3s$^2$3p $^2$P$^o$ and 3s3p$^2$ $^4$P$^e$, both theoretically
whereas only the former experimentally. Synchrotron radiation measurement [8] and 
photoion spectroscopy [13] give its position at 175.25 eV and 175.165$\pm$0.050 eV 
respectively. Theoretically, a 570-term 25 angular component wave function gives an 
energy of $-1.043414$ a.u., and the position at 174.11 eV, in an SP calculation with 
R-matrix approximation [13], which agrees quite well with the recent CCR result of 
$-1.043$ a.u., involving correlated basis sets [53], and the state specific result [55] 
of $-$1.0409856 a.u., as well as with the the multiconfiguration Dirac-Fock [8] excitation 
energy of 174.14 eV. Our energy value of $-1.01210$ a.u., gives its position at 
175.940 eV, which is about 0.67 eV above the experimental value of [8] and matches well 
with the state specific result of 175.15 eV [55]. Our calculated state energy of 
$-$1.02288 a.u., for the latter matches closely with the state specific result of 
$-$1.0393859 a.u. [55]. No other results are available for any of these states for 
further comparison and these results may be useful in future studies of these 
resonances. It may be noted that the present result gives 3s3p$^2$ $^4$P$^e$ as the 
lowest n=3 resonance rather than the 3s$^2$3p $^2$P$^o$, the former lying 0.0108 a.u., 
below the latter. This ordering is in keeping with the CCR findings of [56] for
He$^-$ as well as that of the CI calculation for N$^{4+}$ [22]. However it differs 
from that of [55], who find in a large scale state specific calculation that as in the
n=2 resonances, the 3s$^2$3p $^2$P$^o$ lies below 3s3p$^2$ $^4$P$^e$ for Li 
isoelectronic series, mainly because of the localized electron correlation; the 
separation for Li being about 0.0016 a.u. Clearly, better correlation functionals 
would be required to achieve such smaller separations (of the order of 
1$\times$10$^{-3}$ a.u.) within this DFT formalism. Now Fig. 1 depicts the radial 
densities for some representative (a) $2l2l'$n$l''$ and (b) $3l3l'$n$l''$ hollow 
states and show the characteristic shell structures (superpositions of the orbital
radial densities).

\begingroup
\squeezetable
\begin{table}
\caption {\label{tab:table7}Calculated term energies (in a.u.) and excitation 
energies (in eV) of Li relative to the ground state [54], of some selected 
$3l3l'$n$l''$ states.}
\begin{ruledtabular}
\begin{tabular}{lllllllll}
State & $-$E & Exc. energy & State & $-$E & Exc. energy & 
State & $-$E & Exc. energy \\  
\hline
3s$^2$4s  $^2$S$^e$   &  0.90054  &  178.959   &
3p$^3$  $^4$S$^o$     &  1.00055  &  176.238   & 
3p$^2$4s  $^4$P$^e$   &  0.89860  &  179.012   \\ 
3s$^2$5s  $^2$S$^e$   &  0.87011  &  179.787   &
3p$^3$  $^2$D$^o$     &  0.96847  &  177.111   & 
3p$^2$5s  $^4$P$^e$   &  0.86978  &  179.796   \\ 
3s$^2$6s  $^2$S$^e$   &  0.85729  &  180.136   &
3s3p4s  $^4$P$^o$     &  0.93313  &  178.072   & 
3p$^2$6s  $^4$P$^e$   &  0.85744  &  180.132   \\ 
3s$^2$3p  $^2$P$^o$   &  1.01210\footnotemark[1]  &  
175.924\footnotemark[2]$^,$\footnotemark[3]   &
3s3p5s  $^4$P$^o$     &  0.90193  &  178.921   & 
3p$^2$4s  $^2$D$^e$   &  0.87282  &  179.713   \\ 
3s$^2$4p  $^2$P$^o$   &  0.89129  &  179.211   &
3s3p6s  $^4$P$^o$     &  0.88901  &  179.273   & 
3p$^2$5s  $^2$D$^e$   &  0.84534  &  180.461   \\ 
3s$^2$5p  $^2$P$^o$   &  0.86656  &  179.883   &
3s3p4p  $^4$D$^e$     &  0.92678  &  178.245   & 
3p$^2$6s  $^2$D$^e$   &  0.83330  &  180.788   \\ 
3s$^2$6p  $^2$P$^o$   &  0.85558  &  180.182   &
3s3p5p  $^4$D$^e$     &  0.89924  &  178.994   & 
3p$^2$4p  $^4$D$^o$   &  0.89642  &  179.071   \\ 
3s$^2$3d  $^2$D$^e$   &  0.97108  &  177.040   & 
3s3p6p  $^4$D$^e$     &  0.88767  &  179.309   & 
3p$^2$5p  $^4$D$^o$   &  0.86846  &  179.832   \\ 
3s$^2$4d  $^2$D$^e$   &  0.88088  &  179.494   &
3s3p3d  $^4$F$^o$     &  1.02643  &  175.534   & 
3p$^2$6p  $^4$D$^o$   &  0.85669  &  180.152   \\ 
3s$^2$5d  $^2$D$^e$   &  0.86240  &  179.997   &
3s3p4d  $^4$F$^o$     &  0.91125  &  178.667   & 
3p$^2$4p  $^2$F$^o$   &  0.86932  &  179.808   \\ 
3s$^2$6d  $^2$D$^e$   &  0.85352  &  180.238   & 
3s3p5d  $^4$F$^o$     &  0.89400  &  179.137   & 
3p$^2$5p  $^2$F$^o$   &  0.84343  &  180.513   \\ 
3s3p$^2$  $^4$P$^e$   &  1.02288\footnotemark[4]  &  175.630   & 
3s3p6d  $^4$F$^o$     &  0.88517  &  179.377   & 
3p$^2$6p  $^2$F$^o$   &  0.83225  &  180.817   \\ 
3s3p$^2$  $^2$D$^e$   &  0.99479  &  176.394   & 
                  &               &            & 
                  &               &            \\ 
\end{tabular}
\end{ruledtabular}
\footnotetext[1]{Reference theoretical values are:  1.043414 a.u. [13], 1.043 a.u. 
[53] and 1.040985 a.u. [55].}
\footnotetext[2]{Reference experimental results are: 175.25 eV [8] and 
175.165$\pm$0.050 eV [13].} 
\footnotetext[3]{Reference theoretical values are: 174.11 eV [13], 174.14 eV [8] and
175.15 eV [55].}
\footnotetext[4]{Reference theoretical value is: 1.0393859 a.u. [55].}
\end{table}
\endgroup

\begin{figure}
\begin{minipage}[c]{0.40\textwidth}
\centering
\includegraphics[scale=0.45]{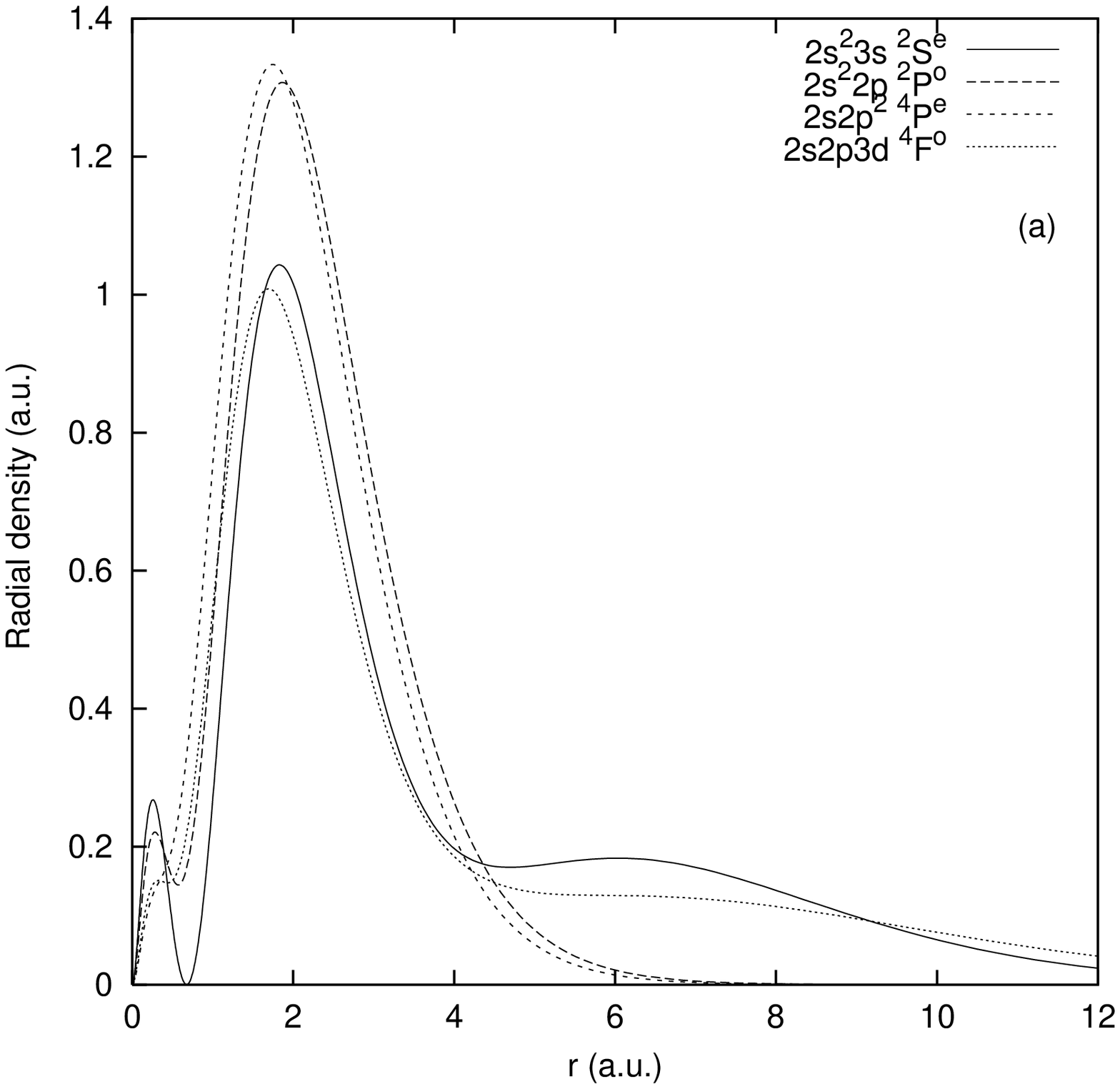}
\end{minipage}%
\hspace{1in}
\begin{minipage}[c]{0.40\textwidth}
\centering
\includegraphics[scale=0.45]{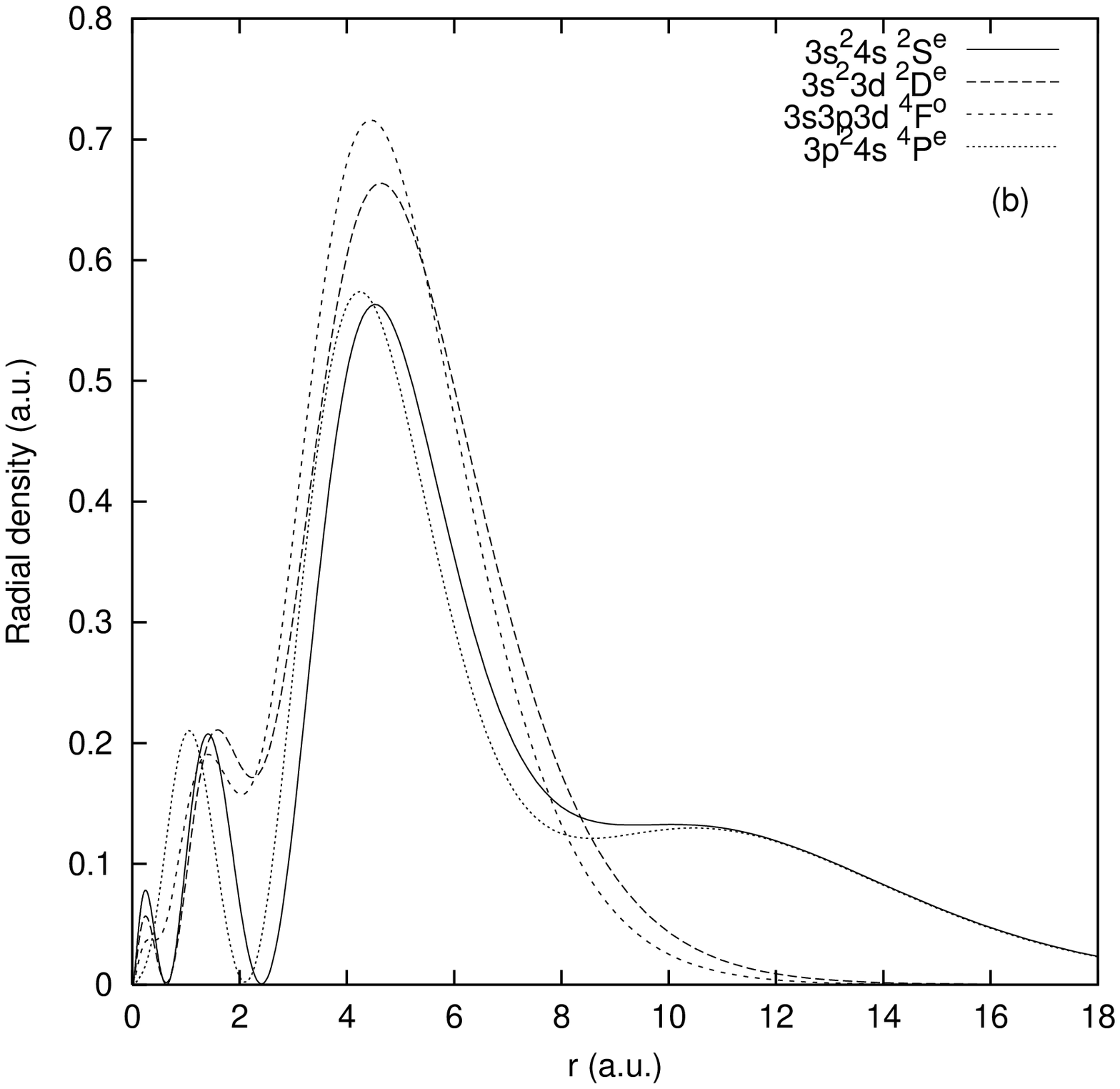}
\end{minipage}%

\caption{The radial densities (a.u.) of Li for (a) 2s$^2$3s $^2$S$^e$, 2s$^2$2p 
$^2$P$^o$, 2s2p$^2$ $^4$P$^e$, 2s2p3d $^4$F$^o$ and (b) 3s$^2$4s $^2$S$^e$, 3s$^2$3d 
$^2$D$^e$, 3s3p3d $^4$F$^o$, 3p$^2$4s $^4$P$^e$ states respectively.}
\end{figure}

At this stage it is worthwhile to make a few pertinent remarks on the status of 
excited state calculations within DFT. Although founded in the 1920s and later
rejuvenated in the 1960s in the works of Hohenberg, Kohn and Sham [58,59], DFT has 
been a very powerful and successful tool for the electronic structure calculation of 
atoms, molecules, solids in their ground states [60,61], inherent difficulties were 
encountered for the excited states and consequently in areas such as spectroscopy, 
its success has been relatively less conspicuous. The well-known fundamental problems 
such as the lack of a unique exact formal relationship valid for general excited states
parallel to the KS method for ground states, as well as the unavailability of hitherto
unfound universal XC energy density functional etc., have been well documented in the 
literature in considerable length (see, for example, [43] for a review). Nevertheless, 
numerous attractive and elegant formalisms have been suggested by many authors over 
the years; e.g., the subspace formulation of DFT [62] and its application to atomic 
excited states [63], ensemble formalism for the unequally weighted states using a 
Rayleigh-Ritz variational principle [64], a perturbative treatment [65,66], calculation 
of multiplet energies within the Hartree-Fock-Slater method utilizing the 
spin-polarized form of the density matrix and exploiting Slater's sum rule [52,67,68], 
ensemble approach using several appropriate functionals [69,70], the time-dependent 
(TD) DFT formulation [71-73] making use of the TD DF response theory, etc. Recently 
correlation energies of several atomic excited states were calculated using the MCSCF 
wave functions [74]. While some of these methods offer good quality results, others 
produce large errors and yet others are computationally difficult to implement. 
Moreover most of these methods have dealt with the lower and singly excited states; 
multiple and higher excitations, especially the Rydberg series such as the ones 
studied in this work, have not been reported so far using any other DFT approach
except the work-function formalism. Besides, while some of these methods, such as 
TDDFT, provides a way to obtain excitation energies in an efficient and accurate 
manner (calculates the linear response of the system to a TD perturbation, leading to 
frequency dependent dynamic dipole polarizability, whose poles and residues yield the 
excitation energies and oscillator strengths respectively), the extraction of 
individual state energies as well as the densities, are not straightforward. In the 
present method however, the energies, excitation energies as well the densities and 
expectation values are obtained easily with reasonable accuracy. To the best of our 
knowledge, this is the first report of triply excited Rydberg resonances of 
many-electron systems within a DFT-based formalism. We also note that the spin 
polarized version of the work-function exchange potential can be obtained as a further 
approximation to the accurate spin polarized X-only KS potential (obtained from a 
consideration of the optimized effective potential method) [75]. Various interesting 
features of the method may be found in the references [27,38-44].

\section{Concluding Remarks}
Twelve triply excited $2l2l'$n$l''$ hollow Rydberg series of Li (covering a total of
about 270 states) have been studied using a density functional approach. 
Nonrelativistic term energies, excitation energies and radial densities are reported 
for both low as well as higher members (up to n=25) of these odd- and even-parity 
resonances. The GPS solution of the KS type equation yields results in fairly good 
agreement with the existing literature data. Additionally 37 $3l3l'$n$l''$ hollow 
states are also presented in the photon energy range 175.63--180.51 eV, out of which 
only two (3s$^2$3p $^2$P$^o$ and 3s3p$^2$ $^4$P$^e$) have been reported so far in the 
literature. The resonance series interact with each other resulting in very 
complicated behavior in their positions. Many new resonances are reported here for the
first time. 

The combination of work-function exchange and the LYP correlation functional within
an essentially single determinantal framework offers results quite comparable in 
accuracy with those from other sophisticated and elaborate quantum mechanical methods.
The exchange-only results are practically of HF-quality; a feature of the work-function
formalism observed in a number of physical systems including atoms, ions, metals, etc.,
(see, for example [76,77]). Consequently, since these are strongly correlated systems, 
one of the main sources of error in this calculation could be due to the inefficiency of
the LYP functional to incorporate the subtle and intricate correlation 
effects, which may be either improved or replaced by more accurate density 
functionals. However, this may not be misconstrued as a drawback of the methodology; 
in fact the results are rather quite encouraging, especially in the light of DFT's 
apparent weaknesses and lack of any density-based attempts as yet for these Rydberg 
series. The assumption of spherical symmetry in calculating the exchange potential, 
might be the another possible cause of inaccuracy. Thus the rotational component of 
the electric field may not have negligible contribution compared to the irrotational 
component for these states, although for atoms this usually holds true [49]. The 
extension of this prescription to even higher photon-energy hollow states such as the 
case where all three electrons remain in shells with n$\ge$4, i.e., the KLM vacancy 
states etc., are straightforward, as well as its application to positive and negative 
ionic systems and extension to the relativistic domain. It may also be interesting to 
employ some of the other DFT-based approaches to treat these and other similar systems, 
so that the nature of the intricate electron correlation may be understood better. Some 
works in these directions are under progress. 

\begin{acknowledgments}
I gratefully acknowledge the warm hospitality provided by the University of New 
Brunswick, Fredericton, Canada. I thank the two anonymous referees for their 
constructive comments. 
\end{acknowledgments}

\end{document}